\documentclass[reqno,11pt]{amsart}
\usepackage{graphicx}
\usepackage{slashed}
\usepackage{amscd}
\usepackage{amssymb}
\usepackage{esint}
\usepackage{enumitem}
\usepackage[mathscr]{eucal}
\numberwithin{equation}{section}
\allowdisplaybreaks[1]

\title[Causal Fermion Systems as a Candidate for a Unified Physical Theory]{Causal Fermion Systems as a Candidate \\ for a Unified Physical Theory}

\author[F.\ Finster]{Felix Finster}
\author[J.\ Kleiner]{Johannes Kleiner \\ \\ February 2015}
\address{Fakult\"at f\"ur Mathematik \\ Universit\"at Regensburg \\ D-93040 Regensburg \\ Germany}
\email{finster@ur.de, Johannes.Kleiner@ur.de}

\newtheorem{Def}{Definition}[section]

\newcommand{\Thanks}{\vspace*{.5em} \noindent \thanks}
\newcommand{\beq}{\begin{equation}}
\newcommand{\eeq}{\end{equation}}
\newcommand{\Proof}{\begin{proof}}
\newcommand{\QED}{\end{proof} \noindent}

\newcommand{\la}{\langle}
\newcommand{\ra}{\rangle}

\newcommand{\Sl}{\mbox{$\prec \!\!$ \nolinebreak}}
\newcommand{\Sr}{\mbox{\nolinebreak $\succ$}}

\newcommand{\C}{\mathbb{C}}
\newcommand{\R}{\mathbb{R}}
\newcommand{\1}{\mbox{\rm 1 \hspace{-1.05 em} 1}}
\newcommand{\Z}{\mathbb{Z}}
\newcommand{\N}{\mathbb{N}}
\newcommand{\Pdd}{\mbox{$\partial$ \hspace{-1.2 em} $/$}}
\newcommand{\Aslsh}{\slashed{A}}

\DeclareMathOperator{\Tr}{Tr}
\DeclareMathOperator{\tr}{tr}

\renewcommand{\L}{{\mathcal{L}}}
\newcommand{\Sact}{{\mathcal{S}}}
\newcommand{\T}{{\mathcal{T}}}

\newcommand\B{{\mathscr{B}}}
\newcommand{\U}{\text{\rm{U}}}
\newcommand{\SU}{\text{\rm{SU}}}

\DeclareMathOperator{\supp}{supp}
\renewcommand{\H}{\mathscr{H}}

\newcommand{\Lin}{\text{\rm{L}}}
\newcommand{\F}{{\mathscr{F}}}

\newcommand{\scrM}{\mycal M}

\newcommand{\itemD}{\item[{\raisebox{0.125em}{\tiny $\blacktriangleright$}}]}
\newcommand{\x}{{\textit{\myfont x}}}
\newcommand{\y}{{\textit{\myfont y}}}
\newcommand*{\myfont}{\fontfamily{ppl}\selectfont}		%ppl
\newcommand{\UMNS}{U_\text{\tiny{MNS}}}

\DeclareFontFamily{OT1}{rsfso}{}
\DeclareFontShape{OT1}{rsfso}{m}{n}{ <-7> rsfso5 <7-10> rsfso7 <10-> rsfso10}{}
\DeclareMathAlphabet{\mycal}{OT1}{rsfso}{m}{n}

\begin{document}

\maketitle

\begin{abstract}
The theory of causal fermion systems is an approach to describe fundamental physics.
Giving quantum mechanics, general relativity and quantum field theory as limiting cases, it is  a candidate for a unified physical theory. We here give a non-technical introduction.
\end{abstract}

\tableofcontents

This article is an introduction to causal fermion systems which is intended to explain
the basic concepts and the general physical picture behind the approach.
The article is organized as follows. In Section~\ref{Theory} we define the basic objects of the theory.
In Section~\ref{Intro-Minkvac} we proceed
by explaining how those objects appear naturally in the familiar physical situation of Dirac particles
in Minkowski space. In Section~\ref{secinherent} it is shown how the objects of quantum
mechanics are encoded in a causal fermion system.
Section~\ref{secminkvac} explains in the example of the Minkowski vacuum how the
causal fermion system encodes the causal structure.
In Section~\ref{secgenst} we exemplify how to describe other physical situations or more general space-times.
In Section~\ref{Continuum} we outline a limiting case in which the causal fermion system
can be described by a second-quantized Dirac field coupled to classical gauge fields and gravity.
Section~\ref{Foundations} gives the resulting perspective on the foundations of quantum mechanics,
in particular on the measurement problem. In Section~\ref{ClarifyingRemarks} we conclude
with a few clarifying remarks.

\section{The Theory} \label{Theory}
The general structure of the theory of causal fermion systems can be understood in analogy to
general relativity. In general relativity, our universe is described by a four-dimensional
space-time (Lorentzian manifold) together with particles and fields.
However, not every configuration of Lorentzian metric, particles and fields is
considered to be ``physical'' in the sense that it could be realized in nature.
Namely, for the configuration to be physically realizable, the Einstein equations
must hold.
Moreover, the particles must satisfy the equations of motion, and the additional fields
must obey the field equations (like Maxwell's equations).
This means that in general relativity, there are two conceptual parts: on the one hand one has
mathematical objects describing possible configurations, and on the other hand there is
a principle which singles out the physical configurations. \\[-0.75em]

The theory of causal fermion systems has the same conceptual structure
consisting of mathematical objects and a principle which singles out the physical configurations.
We first introduce the mathematical objects:
\begin{Def}  {\em{ \textbf{(Causal fermion system)} \label{CFS}
\begin{itemize}[leftmargin=2em]
\itemD Let $(\H,\langle.|.\rangle_\H)$ be a separable complex Hilbert space.
\itemD Given a parameter~$n \in \N$ (the {\em{spin dimension}}), let $\F \subset \Lin(\H)$ be the set of all self-adjoint operators on $\H$ of finite rank, which (counting multiplicities) have at most~$n$ positive and
at most~$n$ negative eigenvalues.
\itemD Let $\rho$ be a positive measure on $\F$ (the {\em{universal measure}}).
\end{itemize}
Then $(\H,\F,\rho)$ is a \textit{causal fermion system}. }}
\end{Def} \noindent
Here separable means that the Hilbert space has an at most countable orthonormal basis.
Mapping the basis vectors to each other, one sees that any two Hilbert spaces are isomorphic,
provided that their dimensions coincide. Therefore, the structure~$(\H, \F)$
is completely determined by the parameters~$n \in \N$ and~$f:= \dim \H \in \N \cup \{\infty\}$.
Apart from these parameters, the only object specifying a causal fermion system is the
universal measure~$\rho$.

It will be outlined below that this definition indeed generalizes mathematical structures used in contemporary physics. The picture is that one causal fermion system describes a space-time together with all structures and objects therein (including the metric, particles and fields). \\[-0.75em]

Next, we state the principle which singles out the physical configurations.
Similar to the Lagrangian formulation of contemporary physics, we work with a
variational principle, referred to as the {\em{causal action principle}}. It states that
a causal fermion system which can be realized in nature should be
a minimizer of the so-called causal action.
In order to formulate the causal action principle, we assume that the
Hilbert space $(\H,\langle.|.\rangle_\H)$ and the spin dimension $n$ have been chosen.
Let $\F$ be as in Definition~\ref{CFS} above. 
Then for any~$x, y \in \F$, the product~$x y$ is an operator
of rank at most~$2n$. We denote its non-trivial eigenvalues (counting algebraic multiplicities)
by~$\lambda^{xy}_1, \ldots, \lambda^{xy}_{2n} \in \C$.
We introduce the {\em{spectral weight}}~$| \,.\, |$ of an operator as the sum of the absolute values
of its eigenvalues. In particular, the spectral weight of the operator
products~$xy$ and~$(xy)^2$ is defined by
\[ |xy| = \sum_{i=1}^{2n} \big|\lambda^{xy}_i \big|
\qquad \text{and} \qquad \left| (xy)^2 \right| = \sum_{i=1}^{2n} \big| \lambda^{xy}_i \big|^2 \:. \]
Next, the {\em{Lagrangian}} $\L : \F \times \F \rightarrow \R_0^+$ is defined by
\begin{align}
\label{Lagr}
\L(x,y) := \big|(xy)^2 \big| - \frac{1}{2n} |xy|^2 = \frac{1}{4n} \sum _{i,j=1}^{2n}
\Big( \big| \lambda_i^{xy} \big| - \big| \lambda_j^{xy} \big| \Big)^2 \:.
\end{align}
The particular form of this Lagrangian is the result of research carried
out over several years (see Section~\ref{seccap}).

\begin{Def}  {\em{ \label{CVP} \textbf{(Causal action principle)}
The {\em{causal action}}~$\Sact$ is obtained by integrating the Lagrangian with respect to the universal measure,
\[ \Sact(\rho) = \iint_{\F \times\F} \L(x,y)\: d\rho(x) \: d\rho(y) \:. \]
The {\em{causal action principle}} is to minimize~$\Sact$ under variations of the universal measure,
taking into account the following constraints:
\begin{align}
&\text{\em{volume constraint:}} & \rho(\F) = \text{const} \quad\;\; & \label{volconstraint} \\
&\text{\em{trace constraint:}} &  \int_\F \tr(x)\: d\rho(x) = \text{const}& \label{trconstraint} \\
&\text{\em{boundedness constraint:}} &   \T := \iint_{\F \times \F} |xy|^2\: d\rho(x)\, d\rho(y) &\leq C \:,
\end{align}
where~$C$ is a given constant (and~$\tr$ denotes the trace of a linear operator on~$\H$). }}
\end{Def}

In mathematical terms, the measure~$\rho$ is varied within the class of
positive regular Borel measures on~$\F$, where on~$\F$ one takes the topology
induced by the $\sup$-norm on~$\Lin(\H)$ (for basic definitions see for example~\cite[Chapters~2 and~5]{rudin}
or~\cite[Chapter~X]{halmosmt}).
The volume and trace constraints are needed in order to avoid trivial minimizers and are important for the analysis of the corresponding Euler-Lagrange equations because they give rise to Lagrange multiplier terms.
The boundedness constraint is needed in order to ensure the existence of minimizers.
In most applications, it does not give rise to a Lagrange multiplier term.
Therefore, it does not seem to have any physical consequences. \\[-0.75em]

This concludes the mathematical definition of the theory.
In order to obtain a physical theory, we need to give the mathematical objects a physical
interpretation. It is one of the main objectives of the next sections to do so by explaining 
how the above mathematical objects relate to the common notions in physics.
The conclusion will be that
causal fermion systems are indeed a candidate for a fundamental physical theory.

\section{Example: Dirac Wave Functions in Minkowski Space}\label{Intro-Minkvac}
As a first step towards explaining how causal fermion systems relate to contemporary physics, we
now explain how the familiar physical situation of Dirac particles in Minkowski space can be
described by a causal fermion system.

Let~$\scrM$ be Minkowski space and $\mu$ the natural volume measure thereon, i.e.~$d\mu = d^4\x$ if $\x=(\x^0,\x^1,\x^2,\x^3)$ is an inertial frame (we use the signature convention $(+,-,-,-)$).
We consider a finite number of~$f$ Dirac particles described by one-particle wave
functions~$\psi_1, \ldots, \psi_f$ which are solutions of the Dirac equation,
\begin{align}
\label{DiracEq}
\big( i \gamma ^j \partial_j - m \big)\, \psi_k = 0, \qquad k=1,\ldots, f\:,
\end{align}
where~$m$ is the rest mass, and~$\gamma^j$ are Dirac matrices in the Dirac representation.
For simplicity, we assume that the wave functions~$\psi_1, \ldots, \psi_f$ are continuous.

Before going on, we remark that this description of the $f$-particle system
by $f$ one-particle wave functions departs from the usual Fock space description.
The connection to Fock spaces will be established later in this article (see
Section~\ref{Foundations}).
For the moment, it is preferable to work with the one-particle wave functions.
We also remark that the assumption of considering a finite number of continuous wave functions
merely is a technical simplification for our presentation. All constructions can be extended
to an infinite number of possibly discontinuous wave functions
(for details see~\cite[Section~4]{finite} or~\cite[Chapter~1]{cfs}).

The wave functions $\psi_k$ span a vector space which we denote by $\H$.
\beq \label{Hspan}
\H := \textrm{span}(\psi_1, \ldots,\psi_f) \:.
\eeq
On~$\H$ we consider the usual scalar product on solutions of the Dirac equation
\begin{align}
\label{ScalProd}
\la \psi | \phi \ra_\H := 2 \pi \int_{t=\textrm{const}} (\overline \psi \gamma^0  \phi) (t,\vec x) \:d^3 x
\end{align}
(here~$\overline{\psi} = \psi^\dagger \gamma^0$ is the adjoint spinor, where the dagger denotes complex conjugation and transposition). If one evaluates~\eqref{ScalProd} for~$\phi=\psi$,
the integrand can be written as~$(\overline{\psi}\gamma^0\psi)(t,\vec{x}) = (\psi^\dagger \psi)(t,\vec{x})$,
having the interpretation as the probability density of the Dirac particle corresponding to~$\psi$
to be at the position~$\vec{x}$. In view of the conservation of probability (being a consequence
of current conservation), the integral in~\eqref{ScalProd} is time independent.
Since the probability density is positive, the inner product~\eqref{ScalProd} is indeed positive definite.
We thus obtain an $f$-dimensional Hilbert space~$(\H, \la .|. \ra_\H)$.

For any~$\x \in \scrM$, we now introduce the sesquilinear form
\[ b_\x :  \H \times \H \rightarrow \C \:,\qquad b_\x(\psi, \phi) = -(\overline \psi \phi) (\x) \:, \]
which maps two solutions of the Dirac equation to their inner product at $\x$.
The sesquilinear form $b_\x$ can be represented by a self-adjoint operator $F(\x)$ on $\H$,
which is uniquely defined by the relations
\[ \la \psi | F(\x) \phi \ra_\H =b_\x(\psi,\phi) \qquad \text{for all~$\psi, \phi \in \H$}\:. \]
More concretely, in the basis~$(\psi_k)_{k = 1, \ldots,f}$ of~$\H$, the last relation can be written as
\begin{align} \label{Fdef}
\la \psi_i | F(\x) \psi_j \ra_\H = - \big(\overline{\psi_i} \psi_j \big)(\x) \:.
\end{align}
If the basis is orthonormal, the calculation
\[ F(\x) \,\psi_j = \sum_{i=1}^f \la \psi_i | F(\x) \psi_j \ra_\H\; \psi_i
= - \sum_{i=1}^f \big(\overline{\psi_i} \psi_j \big)(\x)\; \psi_i \]
(where we used the completeness relation~$\phi = \sum_i \la \psi_i | \phi \ra\, \psi_i$),
shows that the operator~$F(\x)$ has the matrix representation
\[ \big(F(\x) \big)^i_j = - \big(\overline{\psi_i} \psi_j \big)(\x) \:. \]
In physical terms, the matrix element~$-(\overline{\psi_i} \psi_j)(\x)$ gives information on the correlation of the
wave functions~$\psi_i$ and~$\psi_j$ at the space-time point~$\x$.
Therefore, we refer to~$F(\x)$ as the {\em{local correlation operator}} at~$\x$.

Let us analyze the properties of $F(\x)$. First of all, the calculation
\[ \la F(\x) \,\psi \,|\, \phi \ra_\H = \overline{ \la \phi \,|\, F(\x) \,\psi \,\ra_\H}
= -\overline{(\overline \phi \psi) (\x)} = -(\overline \psi \phi) (\x) = \la \psi \,|\, F(\x) \,\phi \ra_\H \]
shows that the operator~$F(\x)$ is self-adjoint
(where we denoted complex conjugation by a bar).
Furthermore, since the pointwise inner product $(\overline \psi \phi)(\x)$ has signature $(2,2)$,
we know that~$b_\x$ has signature $(p,q)$ with $p,q \leq 2$.
As a consequence, the operator $F(\x)$ has at most two positive and at most two negative eigenvalues
(counting multiplicities). It follows immediately, that~$F(\x) \in \F$ if the spin dimension
in Definition~\ref{CVP} is chosen as~$n=2$.

Constructing the operator $F(\x) \in \F$ for every space-time point $\x \in M$, we
obtain the mapping
\begin{align*}
F: \: & \scrM \rightarrow \F \:,\qquad \x \mapsto F(\x) \:.
\end{align*}
This allows us to introduce a measure $\rho$ on $\F$ as follows. For any~$\Omega \in \F$,
one takes the pre-image $F^{-1}(\Omega) \subset \scrM$ and computes its space-time volume,
\[ \rho(\Omega) := \mu \big( F^{-1}(\Omega) \big) \:. \]
This gives rise to the so-called {\em{push-forward measure}} which in mathematics
is denoted by~$\rho = F_\ast \mu$
(we remark for the mathematically oriented reader that the $\sigma$-algebra of
$\rho$-measurable sets is defined as all sets~$\Omega \subset \F$ whose pre-image~$F^{-1}(\Omega)$
is $\mu$-measurable).

Putting the above structures together, we obtain a causal fermion system~$(\H, \F, \rho)$
of spin dimension two. Thus we have succeeded in constructing a causal fermion system
starting from a system of Dirac wave functions in Minkowski space.
But it is not obvious how much of the information on the physical system is encoded in the
causal fermion system. In other words, taking the causal fermion system~$(\H, \F, \rho)$
as the starting point, the question is which structures of the original system can be recovered.
For example, is the Minkowski metric still determined?
\label{Intro-Question1}
Is it possible to reconstruct the Dirac wave functions? Precise answers to these questions will be given
in Section~\ref{secinherent} below. In preparation, we now give a few hints.

We first explain what the points of Minkowski space correspond to in our
causal fermion system. Recall that to every space-time point~$\x \in \scrM$ we associated
a linear operator~$F(\x) \in \F$. Hence the space-time points correspond to the subset~$F(\scrM) \subset \F$.
This subset can also be characterized as the set where the measure~$\rho$ is non-zero.
In mathematical terms, this is captured in the notion of the {\em{support}} of the universal measure,
defined as the set of all the points of~$\F$ such that every open neighborhood of this point
has a non-zero measure. Then (for details see~\cite[Chapter~1]{cfs})
\begin{align}
\label{supp-rho}
\supp{\rho} = \overline{ F(\scrM) }\:,
\end{align}
where the bar denotes the closure.
In all situations of physical interest, the mapping~$F$ will be injective and its image closed
(see again~\cite[Chapter~1]{cfs}). Provided that this is the case,
identifying~$\x \in \scrM$ with the corresponding operator~$F(\x) \in \F$
makes it possible to identify Minkowski space with the support of~$\rho$ as a topological space.
Under suitable smoothness and non-degeneracy assumptions, one can identify~$\scrM$
with~$\supp \rho$ even as a differentiable manifold.
We make this identification manifest by using the letter~$x$ for the operator~$F(\x)$.
In order to avoid confusion, we use two different fonts, making it possible for the
reader to distinguish a point~$\x \in \scrM$ of Minkowski space from
the corresponding point~$x \in M := \supp \rho$.
Once the reader has become familiar with our concepts, the different fonts will be unnecessary.

This consideration shows that the topological and differentiable structures of our space-time
are encoded in the causal fermion system.
Clearly, Minkowski space also has metric and causal structures, which we have not yet
addressed. The general idea for recovering these structures is to take operators~$x,y \in \supp \rho$
and to analyze the eigenvalues of the operator product~$x y$.
The eigenvalues of such operator products contain plenty of information, inducing
relations and structures between the space-time points.
This will be explained more concretely in the next section.

\section{Inherent Structures}\label{secinherent} 
Let~$(\H, \F, \rho)$ be a causal fermion system of spin dimension $n$ (see Definition~\ref{CFS}).
We now introduce additional objects which will turn out to generalize familiar notions in physics.
All of these structures are inherent in the sense that
we only use information already encoded in the causal fermion system.

Motivated by the consideration above (see the paragraph before~\eqref{supp-rho}), we define
{\em{space-time}}~$M$ as the support of the universal measure,
\[ M := \text{supp}\, \rho \subset \F \:. \]
On~$M$ we introduce the following notion of causality.
Recall that for~$x, y \in M$, the product $x y$ is an operator of rank at most $2n$.
We again denote its non-trivial eigenvalues (counting algebraic multiplicities)
by~$\lambda^{xy}_1, \ldots, \lambda^{xy}_{2n} \in \C$.

\begin{Def}  {\em{ \textbf{(Causality)} \label{KS}
The space-time points~$x$ and $y$ are defined to be
\begin{itemize}[leftmargin=2em, itemsep=0.1em]
\itemD \textit{spacelike} separated if all the~$\lambda^{xy}_j$ have the same absolute value.
\itemD \textit{timelike} separated if the~$\lambda_{i}^{xy}$ do not all have the same absolute value
and are all real.
\itemD \textit{lightlike} separated if the~$\lambda_{i}^{xy}$ do not all  have the same absolute value
and are not all real.
\end{itemize} }}
\end{Def}
This definition is compatible with the causal action in the following sense.
If the points~$x$ and~$y$ are spacelike separated, then all the~$\lambda^{xy}_j$ have the same absolute value,
so that the Lagrangian~$\L(x,y)$ vanishes according to~\eqref{Lagr}.
In a more physical language, this means that no interaction takes place 
between regions with spacelike separation
(this does not exclude nonlocal correlations and entanglement, as will be
discussed in Section~\ref{Foundations}).
In this way, our setting incorporates a general version of the principle of causality.

The next step is to introduce wave functions. 
The construction is guided by the usual structure of
a Dirac wave function~$\psi$, which to every space-time point~$\x$
associates a spinor~$\psi(\x)$. The latter is a vector in the corresponding spinor space~$S_\x \scrM \simeq \C^4$,
which is endowed with the inner product~$\overline{\psi} \phi$ of signature~$(2,2)$.
In the setting of causal fermion systems, for a space-time point~$x \in M$ we define the
{\em{spin space}}~$S_x \subset \H$ as the image of the operator~$x$,
\[ S_x := x(\H) \:. \]
It is a subspace of~$\H$ of dimension at most~$2n$.
On~$S_x$ we introduce the inner product
\beq \label{Sprod}
\Sl .|. \Sr_x : S_x \times S_x \rightarrow \C \:,\qquad
\Sl u | v \Sr_x := - \la u | x v \ra _ \H \:,
\eeq
referred to as the {\em{spin scalar product}}.
Since~$x$ has at most~$n$ positive and at most~$n$ negative eigenvalues,
the spin scalar product is an indefinite inner product of signature $(p,q)$ with $p,q \leq n$.
A {\em{wave function}}~$\psi$ is defined as a function
which to every~$x \in M$ associates a vector of the corresponding spin space,
\[ \psi \::\: M \rightarrow \H \qquad \text{with} \qquad \psi(x) \in S_x \quad \text{for all~$x \in M$}\:. \]

Clearly, it is not sufficient to define wave functions abstractly, but we need to specify
those wave functions which are realized in the physical system.
Using a familiar physical language, we need to declare which one-particle states are
occupied (for the connection to multi-particle Fock states see Section~\ref{Foundations}).
To this end, to every vector~$u \in \H$ of the Hilbert space we associate a wave function~$\psi^u$
by projecting the vector~$u$ to the spin spaces, i.e.
\beq \label{psiudef}
\psi^u \::\: M \rightarrow \H\:,\qquad \psi^u(x) := \pi_x u \in S_x \:,
\eeq
where~$\pi_x$ is the orthogonal projection in~$\H$ on the subspace~$x(\H) \subset \H$.
We refer to~$\psi^u$ as the {\em{physical wave function}} corresponding to the vector~$u \in \H$.

Finally, we define the {\em{kernel of the fermionic projector}}~$P(x,y)$
for any~$x, y \in M$ by
\beq \label{Pxydef}
P(x,y) = \pi_x \,y|_{S_y} \::\: S_y \rightarrow S_x
\eeq
(where~$|_{S_y}$ denotes the restriction to the subspace~$S_y \subset \H$).
This object is useful for analyzing the relations and structures between space-time points.
In particular, the kernel of the fermionic projector encodes the causal structure
and makes it possible to compute the eigenvalues~$\lambda^{xy}_1, \ldots, \lambda^{xy}_{2n}$
which appear in the Lagrangian~\eqref{Lagr}. In order to see how this comes about,
we first define the {\em{closed chain}} as the product
\beq \label{Axydef}
A_{xy} = P(x,y)\, P(y,x) \::\: S_x \rightarrow S_x\:.
\eeq
Computing powers of the closed chain and using that~$y \pi_y = y$
(because the image and kernel of self-adjoint operators are orthogonal), we obtain
\[ A_{xy} = (\pi_x y)(\pi_y x)|_{S_x} = \pi_x\, yx|_{S_x} \qquad \text{and thus} \qquad
(A_{xy})^p = \pi_x\, (yx)^p|_{S_x} \:. \]
Taking the trace, we obtain for all~$p \in \N$,
\begin{align*}
\Tr_{S_x} \big( (A_{xy})^p \big) &= \Tr_{S_x} \big(\pi_x\, (yx)^p|_{S_x} \big)
= \tr \big(\pi_x\, (yx)^p|_{S_x} \big) \\
&= \tr \big((yx)^p \pi_x \big)
= \tr \big((yx)^p \big) = \tr \big( (xy)^p \big)
\end{align*}
(where~$\tr$ again denotes the trace of a linear operator on~$\H$).
Since the coefficients of the characteristic polynomial of an operator
can be expressed in terms of traces of powers of the corresponding matrix,
we conclude that the
eigenvalues of the closed chain coincide with the non-trivial
eigenvalues~$\lambda^{xy}_1, \ldots, \lambda^{xy}_{2n}$ of the operator~$xy$ in
Definition~\ref{CVP}.
In this way, one can recover the~$\lambda^{xy}_1, \ldots, \lambda^{xy}_{2n}$ as the eigenvalues of 
a $(2n \times 2n)$-matrix.
In particular, the kernel of the fermionic operator encodes the causal structure of~$M$.

The kernel of the fermionic projector is the starting point for constructions which
unveil the geometric structures of a causal fermion system.
More specifically, this kernel gives rise to a spin connection and corresponding curvature.
Moreover, one can introduce tangent spaces endowed with a Lorentzian metric
together with a corresponding metric connection and curvature.
For brevity, we cannot enter these topics here. Instead we refer the interested
reader to~\cite{lqg, topology}, where also questions concerning the topology
of causal fermion systems are treated.
The important point to keep in mind is that all these constructions are tailored
in order to understand the meaning of information contained in the causal
fermion system. No additional input is required.
The system is completely determined by the causal fermion system~$(\H, \F, \rho)$.
In particular, when varying the universal measure in the causal action principle,
one also varies all the derived structures mentioned above.

\section{The Minkowski Vacuum} \label{secminkvac} 
In order to illustrate the above inherent structures, we now return to the example of Dirac particles in Minkowski
space introduced in Section~\ref{Intro-Minkvac}.
In this example, the Hilbert space~$\H$ is spanned by solutions of the Dirac equation.
Thus a vector~$u \in \H$ is a Dirac wave function, which at a point~$\x \in \scrM$ of
Minkowski space takes values in the corresponding spinor space, $u(\x) \in S_\x\scrM$.
On the other hand, in the previous section we introduced the corresponding physical wave function~$\psi^u$,
which at a point~$x = F(\x) \in M \subset \F$ takes values in the corresponding spin space,
$\psi^u(x) \in S_x$. We now show that these objects can be identified.
Indeed, for any~$u, v \in S_x \subset \H$,
\begin{align*}
\Sl \psi^u(x) \,|\, \psi^v(x) \Sr_x &\overset{\eqref{Sprod}}{=} -\la \pi_x u \,|\, x\, \pi_x v \ra_\H
= -\la u \,|\, x\, v \ra_\H = -\la u \,|\, F(\x)\, v \ra_\H
\overset{\eqref{Fdef}}{=} \overline{u(\x)} v(\x)\:.
\end{align*}
This shows that the inner products on~$S_\x\scrM$ and~$S_x$ are compatible.
It implies that, after choosing suitable bases, one can indeed identify~$S_\x\scrM$ with~$S_x$
(for details see~\cite[Section~1.2]{cfs} or~\cite[Section~4]{lqg}).
This identification implies that~$\psi^u(x) = u(\x)$ for all~$u \in \H$ and~$x \in M$
respectively~$\x \in \scrM$.

Next, it is instructive to bring the kernel of the fermionic projector~\eqref{Pxydef} into a more tractable form.
To this end, we choose an orthonormal basis~$u_1, \ldots, u_f$ of~$\H$. Then
for any~$\phi \in S_y$,
\begin{align*}
P(x,y)\, \phi &= \pi_x \,y\, \phi \overset{(\star)}{=} \sum_{\ell=1}^f \big(\pi_x u_\ell\big) \, \la u_\ell |  \,y\, \phi \ra_\H \\
&\!\!\overset{\eqref{Sprod}}{=} -\sum_{\ell=1}^f  \big(\pi_x u_\ell\big) \,  \Sl \pi_y u_\ell | \phi \Sr_y
\overset{\eqref{psiudef}}{=}  -\sum_{\ell=1}^f  \psi^{u_\ell}(x) \,  \Sl \psi^{u_\ell}(y) \,|\, \phi \Sr_y \:,
\end{align*}
where in~$(\star)$ we used the completeness relation. Using the above identifications
of spinors and their inner products, we can write this formula in the shorter form
\beq \label{Pxyuseful}
P(x,y) = -\sum_\ell u_\ell(\x)\, \overline{u_\ell(\y)}\:.
\eeq
This shows that the kernel of the fermionic projector is composed of all the physical wave functions
of the system.

In order to work in a more concrete example, we next consider the {\em{Minkowski vacuum}}. 
To this end, we want to implement the concept of the Dirac
sea which in non-technical terms states that 
in the vacuum all the negative-energy states of the Dirac equation should be occupied
(see Section~\ref{sec73} for further explanations of this point).
In order to implement this concept, one needs to consider an infinite number of
physical wave functions. This can be achieved simply by letting~$\H$ in Definition~\ref{CFS}
be an infinite-dimensional Hilbert space.
However, a difficulty arises in the construction of the local correlation operators,
because the Dirac wave functions (being square-integrable functions) are in general not defined pointwise,
so that the right side of~\eqref{Fdef} is ill-defined.
In order to resolve this problem, one needs to introduce an ultraviolet regularization.
For conceptual clarity, we postpone the explanation of the ultraviolet regularization to Section~\ref{Continuum}
and now merely mention that
an ultraviolet regularization amounts to modifying the Dirac wave functions
on a microscopic scale~$\varepsilon$, which can be thought of as the Planck scale.
In order to avoid the technical issues involved in the regularization,
we here simply use the formula~\eqref{Pxyuseful},
but now sum over all negative-energy solutions of the Dirac equation.
This sum can be rewritten as an integral over the lower mass shell
(see again~\cite[Section~1.2]{cfs}),
\beq \label{Pxyvac}
P(x,y) = \int \frac{d^4k}{(2 \pi)^4}\:(k_j \gamma^j+m)\: \delta(k^2-m^2)\: \Theta(-k_0)\: e^{-ik(\x-\y)} \:.
\eeq
In this formula, the necessity for an ultraviolet regularization is apparent
in the fact that the Fourier integral is not defined pointwise, but only in the distributional sense
More precisely, the distribution~$P(x,y)$ is singular if the vector~$\xi := \y- \x$ is lightlike,
but it is a smooth function otherwise
(as can be verified for example by explicit computation).
As a consequence, a typical ultraviolet regularization will affect the behavior of~$P(x, y)$ only
in a small neighborhood of the light cone of the form~$\big| |\xi^0| - |\vec{\xi}| \big| \lesssim \varepsilon$.
With this in mind, for the following argument we may disregard the ultraviolet regularization
simply by restricting attention to the region outside this neighborhood.

The representation~\eqref{Pxyvac} allows us to understand the relation
between the Definition~\ref{KS} and the usual notion of causality in Minkowski space:
Since the expression~\eqref{Pxyvac} is Lorentz invariant and is composed of a vector
and a scalar component, the function~$P(x, y)$ can be written as
\[ P(x, y) = \alpha\, \xi_j \gamma^j + \beta\:\1 \]
with two complex-valued functions~$\alpha$ and~$\beta$ (where again~$\xi =\y-\x$).
Taking the conjugate with respect to the spin scalar product, we see that
\[ P(y, x) = P(x, y)^* = \overline{\alpha}\, \xi_j \gamma^j + \overline{\beta}\:\1 \:. \]
As a consequence,
\[ A_{xy} = P(x,y)\, P(y,x) = a\, \xi_j \gamma^j + b\, \1 \]
with two real-valued functions~$a$ and $b$ given by
\[ a = \alpha \overline{\beta} + \beta \overline{\alpha} \:,\qquad
b = |\alpha|^2 \,\xi^2 + |\beta|^2 \:. \]
Applying the formula~$(A_{xy} - b \1)^2 = a^2\:\xi^2\,\1$,
the roots of the characteristic polynomial of~$A_{xy}$ are computed by
\[ b \pm \sqrt{a^2\: \xi^2} \:. \]
Thus if the vector~$\xi$ is timelike, the term~$\xi^2$ is positive, so that
the~$\lambda_j$ are all real. By explicit computation one sees that the coefficients~$a$
and~$b$ are non-zero (see~\cite[Section~\S1.2.5]{cfs}), implying that the
eigenvalues~$\lambda_j$ do not all have the same absolute value. Conversely, if the vector~$\xi$
is spacelike, then the term~$\xi^2$ is negative. Thus the~$\lambda_j$ form a complex conjugate
pair, implying that they all have the same absolute value.
We conclude that the notions of spacelike and timelike as defined for causal fermion systems in
Definition~\ref{KS} indeed agree with the usual notions in Minkowski space.
We remark that this simple argument cannot be used for lightlike directions because
in this case the distribution~$P(x,y)$ is singular, making it necessary to consider an
ultraviolet regularization (the reader interested in the technical details is referred to~\cite{reg}).

To summarize, we have seen that the inherent structures of a causal fermion system
give back the usual causal structure if one considers the Dirac sea vacuum in Minkowski space.
Indeed, a more detailed analysis reveals that the additional inherent structures
mentioned at the end of Section~\ref{secinherent} also give back the
geometric structures of Minkowski space (like the metric and the
connection).

\section{Description of More General Space-Times} \label{secgenst}
The constructions explained above also apply to more general physical situations.
First, one can consider systems involving {\em{particles}}
and {\em{anti-particles}} by occupying additional states and removing
states from the Dirac sea, respectively.
Moreover, our construction also apply in {\em{curved space-time}} (see~\cite{finite, infinite})
or in the presence of an {\em{external potential}} (see~\cite{hadamard}). In all these situations, the
resulting causal fermion systems again encode all the information on the physical system (see~\cite{lqg, cfs}).

The framework of causal fermion systems also allows to describe generalized space-times
(sometimes referred as quantum space-times).
We now illustrate this concept in the simple example of a {\em{space-time lattice}}. Thus we replace Minkowski
space by a four-dimensional lattice~$\scrM := (\varepsilon \Z)^4$ of lattice spacing~$\varepsilon$.
Likewise, the volume measure~$d^4\x$ is replaced by a counting measure~$\mu$
(thus~$\mu(\Omega)$ is equal to the number of lattice points contained in~$\Omega$).
Restricting the Dirac spinors of Minkowski space to the lattice, one gets a
spinor space~$S_\x \scrM$ at every point~$\x \in \scrM$. Dirac wave functions~$\psi_1, \ldots, \psi_f$
can again be introduced as mappings which to every~$\x \in \scrM$ associate a
vector in the corresponding spinor space. These Dirac wave functions can be chosen
for example as solutions of a discretized version of the Dirac equation. Again choosing~$\H$
as the span of the wave functions~\eqref{Hspan} and choosing a suitable scalar product~$\la .|. \ra_\H$,
one defines the local correlation operators again by~\eqref{Fdef}.
Introducing the universal measure as the push-forward of the counting measure~$\mu$,
we obtain a causal fermion system~$(\H, \F, \rho)$ of spin dimension two.
The only difference to the causal fermion system in Minkowski space as constructed
in Section~\ref{Intro-Minkvac} is that now the universal measure is not a continuous
but a discrete measure. 

When describing the Dirac sea vacuum on the lattice, the lattice spacing gives rise to
a natural ultraviolet regularization on the scale~$\varepsilon$.
For example, one may consider all plane-wave solutions~$\psi(x) \sim e^{i kx}$ of the Dirac
equation whose four-momenta lie in the first Brillouin zone, i.e.~$-\pi < \varepsilon \,k_j \leq \pi$ for all~$j=0,\ldots, 3$.
Then one introduces~$\H$ as the Hilbert space generated by
all these plane-wave solutions restricted to the lattice.

Other examples of discrete or singular space-times are described in~\cite{topology}.

\section{The Continuum Limit} \label{Continuum} 
In the previous sections we saw that a causal fermion system has inherent structures which
generalize corresponding notions in quantum theory and relativity.
The next task is to analyze the
dynamics of these objects as described by the causal action principle.
To this end, one considers the Euler-Lagrange (EL) equations corresponding to the
causal action. These equations have a mathematical structure
which is quite different from conventional physical equations  (see~\cite{lagrange}). Therefore, the main
difficulty is to reexpress the EL equations in terms of the inherent structures so as to
make them comparable with the equations of contemporary physics.
This can indeed be accomplished in the so-called continuum limit.
Since the mathematical methods needed for the analysis of the continuum limit
go beyond the scope of this paper (for details see~\cite{cfs, PFP}),
here we can only explain the general concept and discuss the obtained results.

We outlined in Sections~\ref{Intro-Minkvac} and~\ref{secminkvac} how
to describe the Minkowski vacuum by a causal fermion system.
Recall that the construction required an {\em{ultraviolet regularization}} on a microscopic scale~$\varepsilon$.
Such a regularization can be performed in many different ways.
The simplest method is to smooth out the wave functions on the microscopic scale by
convolution with a test function. Another method is to introduce a cutoff in momentum space
on the scale~$\varepsilon^{-1}$. Alternatively, one can regularize by putting the system
on a four-dimensional lattice with lattice spacing~$\varepsilon$ (for example as explained
in Section~\ref{secgenst} above).
It is important to note that each regularization gives rise to a different causal fermion system,
describing a physical space-time with a different microstructure.
Thus in the context of causal fermion systems, the regularization has a physical
significance. The freedom in regularizing reflects our lack of knowledge on the
microstructure of physical space-time.
When analyzing the EL equations corresponding to the causal action, it is not obvious why the effective
macroscopic equations should be independent of the regularization details.
Therefore, it is necessary to consider a sufficiently large class of regularizations,
and one needs to analyze carefully how the results depend on the regularization.
This detailed analysis, referred to as the {\em{method of variable regularization}}
(for more explanations see~\cite[\S4.1]{PFP}),
reveals that for a large class of regularizations,
the structure of the effective macroscopic equations is indeed independent of the regularization
(for details see~\cite[Chapters~3-5]{cfs}).

The {\em{continuum limit}} is a method for evaluating the EL equations corresponding
to the causal action in the limit~$\varepsilon \searrow 0$ when the ultraviolet regularization
is removed. The effective equations obtained in this limit can be evaluated conveniently in
a formalism in which the unknown microscopic structure of space-time (as described by the regularization)
enters only in terms of a finite (typically small) number of so-called {\em{regularization parameters}}.

It turns out that the causal fermion system describing the Minkowski vacuum
satisfies the EL equations in the continuum limit (for any choice of the regularization parameters).
If one considers instead a system involving additional particles and anti-particles,
it turns out the EL equations in the continuum limit no longer hold.
In order to again satisfy these equations, we need to introduce an interaction.
In mathematical terms, this means that the universal measure~$\rho$ must be modified.
Expressed in terms of the inherent structures of a causal fermion system, all the physical wave functions
$\psi^{u_k}(x)$ must be changed collectively.
The analysis shows that this collective behavior of all physical wave functions (including the states
of the Dirac sea) can be described by inserting a potential~$\B$ into the
Dirac equation~\eqref{DiracEq},
\beq \label{Cont:DiracEq}
\big( i \Pdd + \B - m \big) \,u_k(\x) = 0, \qquad k=1,\ldots, f
\eeq
(where as usual~$\Pdd = \gamma^j \partial_j$).
Moreover, the EL equations in the continuum limit are satisfied if and only if the potential~$\B$
satisfies field equations. Before specifying these field equations,
we point out that in the above procedure, the
potential~$\B$ merely is a convenient device in order to describe the collective
behavior of all physical wave functions. It should not be considered as a fundamental
object of the theory.
We also note that, in order to describe variations of the physical wave functions,
the potential in~\eqref{Cont:DiracEq} can be chosen arbitrarily. Each choice of~$\B$
describes a different variation of the physical wave functions.
The EL equations in the continuum limit single out the physically admissible potentials
as being those which satisfy the field equations.

In~\cite{cfs} the continuum limit is worked out in several steps beginning from simple systems
and ending with a system realizing the fermion configuration of the standard model.
For each of these systems, the continuum limit gives rise to effective equations for
second-quantized fermion fields
coupled to classical bosonic gauge fields (for the connection to second-quantized bosonic
fields see Section~\ref{secQFT} below).
To explain the structure of the obtained results, it is preferable to first describe the
system modelling the leptons as analyzed in~\cite[Chapter~4]{cfs}.
The input to this model is the configuration of the leptons in the standard model
without interaction. Thus the fermionic projector of the vacuum is assumed to be
composed of three generations of Dirac particles of masses~$m_1, m_2, m_3>0$
(describing~$e$, $\mu$, $\tau$)
as well as three generations of Dirac particles of masses~$\tilde{m}_1, \tilde{m}_2,
\tilde{m}_3 \geq 0$ (describing the corresponding neutrinos).
Furthermore, we assume that the regularization of the neutrinos breaks the chiral
symmetry (implying that we only see their left-handed components).
We point out that the definition of the model does not involve any assumptions on the interaction.

The detailed analysis in~\cite[Chapter~4]{cfs} reveals that the effective interaction in the continuum limit
has the following structure.
The fermions satisfy the Dirac equation coupled to a left-handed $\SU(2)$-gauge
potential~$A_L=\big( A_L^{ij} \big)_{i,j=1,2}$,
\[ \left[ i \Pdd + \begin{pmatrix} \Aslsh_L^{11} & \Aslsh_L^{12}\, \UMNS^* \\[0.2em]
\Aslsh_L^{21}\, \UMNS & -\Aslsh_L^{11} \end{pmatrix} \chi_L
- m Y \right] \!\psi = 0 \:, \]
where we used a block matrix notation (in which the matrix entries are
$3 \times 3$-matrices). Here~$mY$ is a diagonal matrix composed of the fermion masses,
\beq \label{mY}
mY = \text{diag} (\tilde{m}_1, \tilde{m}_2, \tilde{m}_3,\: m_1, m_2, m_3)\:,
\eeq
and~$\UMNS$ is a unitary $3 \times 3$-matrix (taking the role of the
Maki-Nakagawa-Sakata matrix in the standard model).
The gauge potentials~$A_L$ satisfy a classical Yang-Mills-type equation, coupled
to the fermions. More precisely, writing the isospin dependence of the gauge potentials according
to~$A_L = \sum_{\alpha=1}^3 A_L^\alpha \sigma^\alpha$ in terms of Pauli matrices,
we obtain the field equations
\beq \label{l:YM}
\partial^k \partial_l (A^\alpha_L)^l - \Box (A^\alpha_L)^k - M_\alpha^2\, (A^\alpha_L)^k = c_\alpha\,
\overline{\psi} \big( \chi_L \gamma^k \, \sigma^\alpha \big) \psi\:,
\eeq
valid for~$\alpha=1,2,3$ (for notational simplicity, we wrote the Dirac current for one Dirac
particle; for a second-quantized Dirac field, this current is to be replaced by the expectation value
of the corresponding fermionic field operators). Here~$M_\alpha$ are the bosonic masses and~$c_\alpha$
the corresponding coupling constants.
The masses and coupling constants of the two off-diagonal components are
equal, i.e.\ $M_1=M_2$ and~$c_1 = c_2$,
but they may be different from the mass and coupling constant
of the diagonal component~$\alpha=3$. Generally speaking, the mass ratios~$M_1/m_1$, $M_3/m_1$ as
well as the coupling constants~$c_1$, $c_3$ depend on the regularization. For a given regularization,
they are computable.

Finally, our model involves a gravitational field described by the Einstein equations
\beq \label{l:Einstein}
R_{jk} - \frac{1}{2}\:R\: g_{jk} + \Lambda\, g_{jk} = \kappa\, T_{jk} \:,
\eeq
where~$R_{jk}$ denotes the Ricci tensor, $R$ is scalar curvature, and~$T_{jk}$
is the energy-momentum tensor of the Dirac field. Moreover, $\kappa$ and~$\Lambda$ denote the
gravitational and the cosmological constants, respectively.
We find that the gravitational constant scales like~$\kappa \sim \delta^{-2}$, where~$\delta \geq \varepsilon$ is
the length scale on which the chiral symmetry is broken.

In~\cite[Chapter~5]{cfs} a system is analyzed which realizes the
configuration of the leptons and quarks in the standard model.
The result is that the field equation~\eqref{l:YM} is replaced by
field equations for the electroweak and strong interactions after spontaneous
symmetry breaking (the dynamics of the corresponding Higgs field has not yet been analyzed).
Furthermore, the system again involves gravity~\eqref{l:Einstein}.

A few clarifying remarks are in order. First, the above field equations come with
corrections which for brevity we cannot discuss here (see~\cite[Sections~3.8, 4.4 and~4.6]{cfs}).
Next, it is worth noting that, although
the states of the Dirac sea are explicitly taken into account in our analysis, they do not
enter the field equations. More specifically, in a perturbative treatment,
the divergences of the Feynman diagram
describing the vacuum polarization drop out of the EL equations of the causal action.
Similarly, the naive ``infinite negative energy density'' of the
sea drops out of the Einstein equations, making it unnecessary to subtract any counter terms.
We finally remark that the only free parameters of the theory are the masses in~\eqref{mY}
as well as the parameter~$\delta$ which determines the gravitational constant.
The coupling constants, the bosonic masses and the mixing matrices
are functions of the regularization parameters
which are unknown due to our present lack of knowledge on the microscopic structure of space-time.
The regularization parameters cannot be chosen arbitrarily because they must satisfy certain
relations. But except for these constraints, the regularization parameters are currently treated
as free empirical parameters.

To summarize, the dynamics in the continuum limit is described by Dirac spinors
coupled to classical gauge fields and gravity. The effective continuum theory is manifestly covariant
under general coordinate transformations.
The only limitation of the continuum limit is that the bosonic fields are merely classical.
However, as will be briefly mentioned in Section~\ref{secQFT}, a detailed analysis
which goes beyond the continuum limit gives rise even to second-quantized bosonic fields.
Based on these results, the theory of causal fermion systems
seems to be a promising candidate for a unified physical theory.

\section{Connection to Foundations of Quantum Theory}\label{Foundations}
As explained above, in the continuum limit the dynamics of quantum mechanical
wave functions is described by the Dirac equation, which in the non-relativistic limit reduces to the
Pauli equation or the Schr\"odinger equation (cf.~\cite{bjorken, peskin+schroeder}).

It is well-known in the foundations of quantum theory that the linearity of the Schr\"odinger equation
per se seems to be in conflict with the experimental observation of definite
measurement outcomes (one observes a dead cat or an alive cat, but no superposition of the two).
This conflict, referred to as the \textit{measurement problem} (cf.~\cite{peres}), can be remedied in several ways, leading to modifications of the original quantum mechanics as formulated by von Neumann~\cite{vneumann}.
Since the theory of causal fermion systems is a candidate for a unified physical theory, this raises the question
of what its implications are on the measurement problem.
This question is presently under investigation. We here outline how it can
be addressed and formulate a conjecture.

We first point out that the EL equations corresponding to the causal action
are {\em{nonlinear}}. This is obvious in the abstract setting simply because the positive Borel
measures do not form a vector space (the linear combination of two such measures is in general no
longer positive). More explicitly, this nonlinearity can be seen by
reexpressing the EL equations in terms of inherent structures like
the physical wave functions. Namely, according to~\eqref{Pxyuseful}, the fermionic projector
is quadratic in the wave functions. Thus the closed chain~\eqref{Axydef}
is of fourth order in the wave functions. Computing the eigenvalues of~$A_{xy}$
involves square roots, also we need to take absolute values of these eigenvalues.
Therefore, the Lagrangian~\eqref{Lagr} depends on the wave functions in a
highly nonlinear way.
Nevertheless, {\em{linear}} evolution equations are obtained by
considering linear perturbations of the Dirac sea vacuum.
This approximation is justified in many situations because the contribution of,
for example, a single electron wave function to the causal action is very small
compared to the total contribution of all the sea states.
This is the underlying reason why the dynamics of the wave functions as obtained in the
continuum limit is linear. However, this linear dynamics is only an approximation.

Two corrections to the linear Schr\"odinger dynamics seem to be most relevant.
First, a nonlinear correction to the Dirac equation arises by taking into account
the perturbation of the Dirac sea vacuum in the EL equations to second order.
Another effect is due to the microscopic structure of space-time.
Taking into account fluctuations of this microstructure (which are disregarded in
the continuum limit) seems to give rise to a stochastic correction term to the Dirac equation.
This combination of a nonlinear and a stochastic correction term is reminiscent
of the modifications of the Schr\"odinger equation used for example in the spontaneous localization model
(see~\cite{pearle0, ghirardi2, bassi} or~\cite[Chapter~8]{joos}). 
This leads us to the following
\begin{quote} {\em{Conjecture.}}
The theory of causal fermion systems gives rise to an effective dynamical collapse theory.
\end{quote}
By ``dynamical collapse theory'' we mean a Schr\"odinger or Dirac equation involving
nonlinear and stochastic terms which resolve the measurement problem.
``Effective'' means that the nonlinear equations are not taken as the starting point,
but they arise in an effective description
of the dynamics as determined by the EL equations of the causal action.

Other points of interest with respect to foundations of quantum theory are
{\em{nonlocality}} and {\em{entanglement}}. Both are experimentally tested features of quantum theory
which need to be explained by the theory of causal fermion systems.
To understand the role of nonlocality, one should keep in mind that in a causal fermion system,
a fermion is described by a physical wave function~$\psi^u(x)$. 
As in standard quantum mechanics, these wave functions are nonlocal objects
spread out in space-time, giving rise to the usual nonlocal correlations for
one-particle measurements.

In order to describe entanglement, one needs to work with a multi-particle wave function.
The simplest method to obtain the connection is to choose
an orthonormal basis~$u_1, \ldots, u_f$ of~$\H$ and to form the $f$-particle Hartree-Fock state
\beq \label{antisymm}
\Psi := \psi^{u_1} \wedge \cdots \wedge \psi^{u_f} \:.
\eeq
Clearly, the choice of the orthonormal basis is unique only up to the unitary transformations
\[ u_i \rightarrow \tilde{u}_i = \sum_{j=1}^f U_{ij} \,u_j \quad \text{with} \quad U \in \U(f)\:. \]
Due to the anti-symmetrization, this transformation changes the corresponding Hartree-Fock state
only by an irrelevant phase factor,
\[ \psi^{\tilde{u}_1} \wedge \cdots \wedge \psi^{\tilde{u}_f} = \det U \;
\psi^{u_1} \wedge \cdots \wedge \psi^{u_f} \:. \]
Thus the configuration of the physical wave functions can be described by a
fermionic multi-particle wave function.

The shortcoming of the above construction is that the Hartree-Fock state~\eqref{antisymm}
does not allow for the description of entanglement.
But entanglement arises naturally if the effect of {\em{microscopic mixing}} is taken into
account, as we now briefly outline.
Microscopic mixing is based on the observation that the causal action of
a Dirac sea configuration is smaller if the physical wave functions have fluctuations
on the microscopic scale. To be more precise, one constructs a universal measure~$\rho$
which consists of~$L$ components, i.e.\ $\rho = \rho_1 + \cdots +\rho_L$.
This also gives rise to a decomposition of the corresponding space-time, i.e.\
$M= M_1 \cup \cdots \cup M_L$ with~$M_\ell := \supp \rho_\ell$.
Now one considers variations of the measures~$\rho_\ell$ obtained by modifying the
phases of the physical wave functions in the sub-space-times~$M_\ell$.
Minimizing the causal action under such variations, one sees that 
the kernel of the fermionic projector~$P(x,y)$ becomes very small if~$x$ and~$y$
are in different sub-space-times. This effect can be understood similar to
a dephasing of the physical wave functions in different sub-space-times.

The resulting space-time~$M$ has a structure which cannot be understood classically.
One way of visualizing~$M$ is that it consists of
different global space-times~$M_\ell$ which are interconnected by relations between them.
An alternative intuitive picture is to regard~$M$ as a single space-time which is
``fine-grained'' on the microscopic scale by the sub-space-times $M_\ell$.
For the physical wave functions, the above dephasing effect means that
every physical wave function~$\psi^u(x)$ has {\em{fluctuations}} on the microscopic scale.
Moreover, comparing~$\psi^u(x)$ and~$\psi^u(y)$ for~$x$ and~$y$ in the same sub-space-time,
one finds {\em{nonlocal correlations}} on the macroscopic scale.
A detailed analysis shows that taking averages over the sub-space-times
gives rise to an effective description of the interaction in terms of multi-particle wave functions and
Fock spaces (see~\cite[Sections~5, 6 and~8]{qft}).
In particular, this gives agreement with the usual description of entanglement.

To summarize, entanglement arises naturally in the framework of causal
fermion systems when taking into account the effect of microscopic mixing.
The reader who wants to understand the concept of microscopic mixing
on a deeper quantitative level is referred to~\cite{qft}.

\section{Clarifying Remarks}\label{ClarifyingRemarks} 
This section aims to address some of the questions which might have come to the mind
of the reader.

\subsection{Where does the name ``causal fermion system'' come from?} \label{cfsname}
The term {\em{``causality''}} in the name causal fermion system refers to the fact that there are causal
relations among the space-time points (see Definition~\ref{KS}). The causal action is ``causal''
because it vanishes for space-time points with spacelike separation.
In this way, the notion of causality is intimately connected with the framework of causal fermion systems.
The term ``fermion'' refers to the fact that a causal fermion system encodes
physical wave functions~$\psi^u(x)$ (see~\eqref{psiudef}) which are interpreted as
{\em{fermionic}} wave functions (like Dirac waves).
This interpretation as fermionic wave functions is justified because,
rewriting the configuration of the physical wave functions in the
Fock space formalism, one obtains
a totally anti-symmetric multi-particle state (see~\eqref{antisymm}).
{\em{Bosonic}} fields appear in the causal fermion systems merely as a device to describe the collective
behavior of the fermions (see~\eqref{Cont:DiracEq}). But they should not be considered
as fundamental objects of the theory.

\subsection{Why this form of the causal action principle?} \label{seccap}
The first attempts to formulate a variational principle in space-time
in terms of fermionic wave functions can be found in the unpublished preprint~\cite{endlich}.
The variational principle proposed in~\cite[Section~3.5]{PFP} coincides with
the causal action principle, except that it is formulated in the setting of discrete space-times
and that the constraints~\eqref{volconstraint} and~\eqref{trconstraint} are missing.
The general structure of the Lagrangian~\eqref{Lagr} can be understood from the requirements that
it should be non-negative and that it should vanish for spacelike separation.
The detailed form of the Lagrangian~\eqref{Lagr} is determined uniquely by demanding that the
Dirac sea vacuum should be a stable minimizer of the variational principle
(as is made precise by the notion of ``state stability''; see~\cite[Section~5.6]{PFP}).
The necessity and significance of the constraints~\eqref{volconstraint} and~\eqref{trconstraint} became clear
when analyzing the existence theory~\cite{discrete, continuum} and deriving the EL equations~\cite{lagrange}.
It should also be noted that the so-called identity constraint considered in~\cite{continuum}
has turned out to be a too strong condition which is not compatible with the so-called spatial
normalization of the fermionic projector as discussed in~\cite[Section~2.2]{norm} and established
in~\cite{noether}.

\subsection{Why the name ``continuum limit''?}
Causal fermion systems were first analyzed in the more restrictive formulation of discrete space-times
(see~\cite[Section~3.3]{PFP}). In this setting, the continuum limit as introduced in~\cite[Chapter~4]{PFP}
arises as the limit when the discretization scale~$\varepsilon$ tends to zero, meaning that
the discrete space-time goes over to a space-time continuum.
The more general notion of causal fermion systems given here allows for the description
of both continuous and discrete space-times. Then the parameter~$\varepsilon$ should be regarded as
a regularization length, but space-time could very well be continuous on this scale.
In this more general context, the notion ``continuum limit'' merely means that
we take the limit~$\varepsilon \searrow 0$ in which space-time $M:= \supp \rho$ goes over to the
{\em{usual}} space-time continuum~$\scrM$ (i.e.\ Minkowski space or a Lorentzian manifold,
without microscopic mixing).

\subsection{Connection to the notion of the Dirac sea} \label{sec73}
The concept of the Dirac sea was introduced by Dirac in order to remedy the problem of the
negative-energy solutions of the Dirac equation.
Dirac's original conception was that in the vacuum all negative-energy states are occupied.
Due to the Pauli exclusion principle, additional particles must occupy states of positive energy.
This concept led to the prediction of anti-particles, which are described as ``holes'' in the sea.

If taken literally, the concept of the Dirac sea leads to problems such as an infinite negative
energy density or an infinite charge density. This is the main reason why in modern quantum field theory,
the concept of the Dirac sea is no longer apparent. It
has been replaced by Wick ordering and the reinterpretation of creation and annihilation operators
corresponding to the negative-energy states.
Therefore, it is a common view that the Dirac sea is merely a historical relic which is no longer
needed.

In the theory of causal fermion system, Dirac's original concept is revived.
Namely, when constructing a causal fermion system starting from a classical space-time
the states of the Dirac sea need to be taken into account
(cf.~\eqref{Pxyuseful} and~\eqref{Pxyvac} in the Minkowski vacuum).
This can be understood as follows. It is a general concept behind causal fermion systems
that all structures in space-time should be encoded in the physical wave functions.
This concept only works if there are ``sufficiently many'' physical wave functions.
More specifically, this is the case if the causal fermion system is composed of
a regularized Dirac sea configuration, possibly with additional particles and/or anti-particles.

In contrast to the problems in the naive Dirac sea picture, in the description with
causal fermion systems the ensemble of the sea states does {\em{not}} give rise to an infinite negative energy
density or an infinite charge density. Namely, due to the specific form of the causal action principle,
the sea states drop out of the Euler-Lagrange equations in the continuum limit.

\subsection{Connection to quantum field theory} \label{secQFT}
The continuum limit gives an effective description of the interaction on the level of
second-quantized fermionic fields coupled to classical bosonic fields.
A full quantum field theory, in which also the bosonic fields are quantized,
arises if the effect of microscopic mixing is taken into account.
We refer the reader to Section~\ref{Foundations} as well as the article~\cite{qft}.
The detailed analysis of the resulting Feynman diagrams, renormalization
and a comparison with standard quantum field theory is work in progress.

\subsection{Which physical principles are incorporated in a causal fermion system?}
Causal fermion systems evolved from an attempt to combine several physical principles
in a coherent mathematical framework. As a result, these principles appear in the
framework in a specific way:
\begin{itemize}[leftmargin=1.3em, itemsep=0.2em]
\itemD The {\bf{principle of causality}} is built into a causal fermion system in a specific way,
as explained in Section~\ref{cfsname} above.
\itemD The {\bf{Pauli exclusion principle}} is incorporated in a causal fermion system,
as can be seen in various ways. 
One formulation of the Pauli exclusion principle states that every fermionic one-particle state
can be occupied by at most one particle. In this formulation, the Pauli exclusion principle
is respected because every wave function can either be represented in the form~$\psi^u$
(the state is occupied) with~$u \in \H$ or it cannot be represented as a physical wave function
(the state is not occupied). But it is impossible to describe higher occupation numbers.
When working with multi-particle wave functions, the Pauli exclusion principle becomes apparent
in the total anti-symmetrization of the wave function (see~\eqref{antisymm}).
\itemD A {\bf{local gauge principle}} becomes apparent once we choose
basis representations of the spin spaces and write the wave functions in components.
Denoting the signature of~$(S_x, \Sl .|. \Sr_x)$ by~$(p(x),q(x))$, we choose
a pseudo-orthonormal basis~$(\mathfrak{e}_\alpha(x))_{\alpha=1,\ldots, p+q}$ of~$S_x$.
Then a wave function~$\psi$ can be represented as
\[ \psi(x) = \sum_{\alpha=1}^{p+q} \psi^\alpha(x)\: \mathfrak{e}_\alpha(x) \]
with component functions~$\psi^1, \ldots, \psi^{p+q}$.
The freedom in choosing the basis~$(\mathfrak{e}_\alpha)$ is described by the
group~$\U(p,q)$ of unitary transformations with respect to an inner product of signature~$(p,q)$.
This gives rise to the transformations
\[ \mathfrak{e}_\alpha(x) \rightarrow \sum_{\beta=1}^{p+q} U^{-1}(x)^\beta_\alpha\;
\mathfrak{e}_\beta(x) \qquad \text{and} \qquad
\psi^\alpha(x) \rightarrow  \sum_{\beta=1}^{p+q} U(x)^\alpha_\beta\: \psi^\beta(x) \]
with $U \in \U(p,q)$.
As the basis~$(\mathfrak{e}_\alpha)$ can be chosen independently at each space-time point,
one obtains {\em{local gauge transformations}} of the wave functions,
where the gauge group is determined to be the isometry group of the spin scalar product.
The causal action is
{\em{gauge invariant}} in the sense that it does not depend on the choice of spinor bases.
\itemD The {\bf{equivalence principle}} is incorporated in the following general way.
Space-time $M:= \supp \rho$ together with the universal measure~$\rho$ form a topological
measure space, being a more general structure than a Lorentzian manifold.
Therefore, when describing~$M$ by local coordinates, the freedom in choosing such
coordinates generalizes the freedom in choosing general reference frames in a space-time manifold.
Therefore, the equivalence principle of general relativity is respected. The causal action is {\em{generally
covariant}} in the sense that it does not depend on the choice of coordinates.
\end{itemize}

\subsection{Philosophical remarks}
Since causal fermion systems are a candidate for a unified physical theory,
one may take a consistent realist point of view and assume that our universe is a causal fermion
system. Here by ``realist point of view'' we mean that one assumes that there is a
reality independent of human observation and that one can describe this reality in a mathematical
language. ``Consistent'' means that this point of view does not lead to contradictions or
inconsistencies. Finally, by ``universe is a causal fermion system'' we mean that
the fundamental entities of our universe are the causal fermion system~$(\H,\F,\rho)$ as well as its inherent
structures.

This position could be investigated from a philosophical point of view.
We find the following points interesting:
\begin{itemize}[leftmargin=1.3em]
\itemD Space-time is a set of operators. The relations between space-time points
are all encoded in properties of products of these operators.
No additional structures need to be specified.
\itemD Similar to the picture in dynamical collapse theories, the basic object to
describe a fermion is the physical wave function~$\psi^u(x)$.
The particle character, however, comes about merely as a consequence of the dynamics
as described by the causal action principle.
\itemD The structures of space-time and matter are described in terms of a single
object: the universal measure. In particular, it is no longer possible to separate
space-time from the matter content therein.
This seems to go a step further than relativity: In relativity, space and time
do not exists separately, but are combined to space-time.
In the approach of causal fermion systems, space-time does not exist without
the matter content (including the Dirac sea). 
Space-time and the matter content are combined in one object.
\end{itemize}
A further investigation of these and related points might offer new insight on
questions in the philosophy of science.

\Thanks {{\em{Acknowledgments:}}
J.K.\ gratefully acknowledges support by the ``Studienstiftung des deutschen Volkes.''
We would like to thank Meinard Kuhlmann,
Olaf M\"uller, Christian R\"oken and Jan-Hendrik Treude for helpful comments
on the manuscript. We are grateful to Thomas Elze for organizing the inspiring conference in Castiglioncello.

%\bibliographystyle{amsplain}
%\bibliography{../../aarbeit/felix}
\providecommand{\bysame}{\leavevmode\hbox to3em{\hrulefill}\thinspace}
\providecommand{\MR}{\relax\ifhmode\unskip\space\fi MR }
\providecommand{\MRhref}[2]{%
  \href{http://www.ams.org/mathscinet-getitem?mr=#1}{#2}
}
\providecommand{\href}[2]{#2}

\end{document}